\documentclass[twocolumn]{rps-esrl2020}

\def\papername{\jobname}

\usepackage{natbib}

\begin{document}

\markboth{Tagir Fabarisov}{Model-based Fault Injection Experiments for the Safety Analysis of Exoskeleton System}

\twocolumn[

\title{Model-based Fault Injection Experiments for the Safety Analysis of Exoskeleton System}

\author{Tagir Fabarisov}

\address{Institut f{\"u}r Automatisierungstechnik, Technische Universit{\"a}t Dresden, Germany. \email{tagir.fabarisov@tu-dresden.de}}

\author{Ilshat Mamaev}

\address{Karlsruhe Institute of Technology, Germany. \email{ilshat.mamaev@kit.de}}

\author{Andrey Morozov}

\address{University of Stuttgart, Germany. \email{andrey.morozov@ias.uni-stuttgart.de}}

\author{Klaus Janschek}

\address{Institut f{\"u}r Automatisierungstechnik, Technische Universit{\"a}t Dresden, Germany. \email{klaus.janschek@tu-dresden.de}}

\begin{abstract} 
Model-based fault injection methods are widely used for the evaluation of fault tolerance in safety-critical control systems.
In this paper, we introduce a new model-based fault injection method implemented as a highly-customizable Simulink block called FIBlock.
It supports the injection of typical faults of essential heterogeneous components of Cyber-Physical Systems, such as sensors, computing hardware, and network.
The FIBlock GUI allows the user to select a fault type and configure multiple parameters to tune error magnitude, fault activation time, and fault exposure duration.
Additional trigger inputs and outputs of the block enable the modeling of conditional faults. 
Furthermore, two or more FIBlocks connected with these trigger signals can model chained errors.

The proposed fault injection method is demonstrated with a lower-limb EXO-LEGS exoskeleton, an assistive device for the elderly in everyday life.
The EXO-LEGS model-based dynamic control is realized in the Simulink environment and allows easy integration of the aforementioned FIBlocks.
Exoskeletons, in general, being a complex CPS with multiple sensors and actuators, are prone to hardware and software faults.
In the case study, three types of faults were investigated: 1) sensor freeze, 2) stuck-at-0, 3) bit-flip.
The fault injection experiments helped to determine faults that have the most significant effects on the overall system reliability and identify the fine line for the critical fault duration after that the controller could no longer mitigate faults.

\end{abstract}


]

\section{Introduction}

In recent years emerging technologies in human-robot collaboration brought science fiction closer to reality.
Particularly in the domain of wearable robotics, exoskeletons grant users mobility support in the military, medical, and industrial (\citealp{Rewalk,HAL}, \citealp{Robomate,reviewEXOS}) applications.
Being an assistive device exoskeletons are designed to interact with its user and can be controlled by a model, user body motions, sensor feedback, or even mentally via a brain-computer interface.
This close collaboration exposes the human operator to many hazards that require cautious development and the design of exoskeletons.
Several standards guide the inherently safe design of industrial (ISO 10218 /1-2011 and ISO 10218 /2-2011) and personal care robots (ISO 13482 – 2014).
In the latter one, the term "exoskeleton" was first introduced, referring to a physical assistant robot.

EXO-LEGS is an EU-funded (AAL-2011-4) project for developing a prototype for modular commercially viable lower-limb exoskeleton for elderly persons in everyday life (\citealp{EXOLEGS}).
Within this project, an exoskeleton prototype with six active degrees-of-freedom (DoF) was developed (Figure ~\ref{fig:EXO-LEGSCAD}).
The model-based dynamic control with the  Center of Mass (CoM) and the Zero Moment Point (ZMP) stability criteria were realized using MATLAB/Simulink and Hardware-in-the-Loop methods.
Further development of the EXO-LEGS framework integrated human motion capture and dynamic movement primitives generation methods in order to be able to execute recorded motions from everyday life activities like walking, standing up, sitting down, etc.
Such a system has to fulfill high reliability and safety requirements.
In particular, the development process must ensure that the exoskeleton performs safely under the presence of common faults of multiple and heterogeneous components.
This can be achieved via extensive and automated fault injection experiment.

\textbf{Contribution:}
This paper presents a new model-based fault injection method implemented as a highly-customizable Simulink block called FIBlock. 
Our method supports the injection of typical faults of essential heterogeneous components of Cyber-Physical Systems, such as sensors, computing hardware, and network.
The FIBlock GUI allows the user to select a fault type and configure multiple parameters to tune error magnitude, fault activation, and fault exposure times.
Additional trigger inputs and outputs of the block enable the modeling of conditional faults. 
Furthermore, two or more FIBlocks connected with the trigger signals can model so-called chained errors.
The FIBlock is available on the GitHub (\citealp{fabarisov2020fiblock}).
The application is demonstrated on the EXO-LEGS system discussed above.
In this paper we employ presented fault injection method for the fault tolerance assessment of the exoskeleton system.

\section{State of the Art}
\label{sec:sta}

Fault injection helps to evaluate system fault tolerance.
In modern industrial safety standards, the fault injection is a required procedure to verify and ensure system reliability.
The automotive safety standard ISO 26262 requires not only a single fault injection but also a combination of faults.

\subsection{Fault Types}

\begin{figure}[b]
  \centering
  \includegraphics[width=\linewidth]{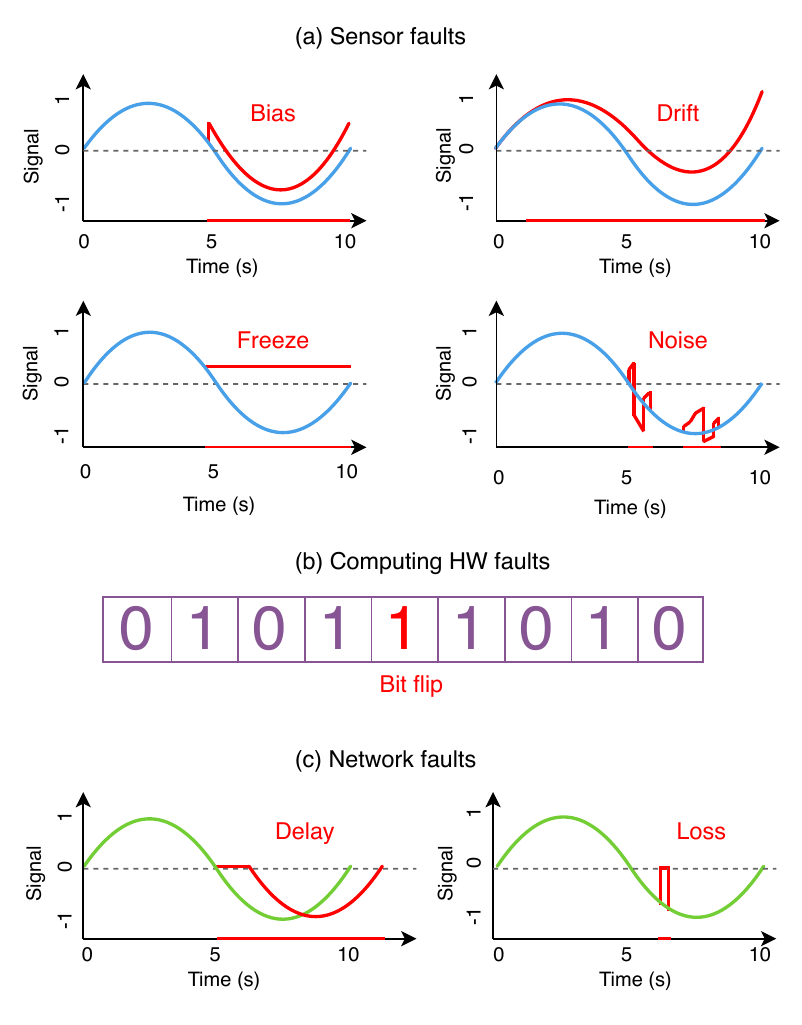}
  \caption{Common fault types of the CPS components (\citealp{ding2019line}).}
  \label{fig:CPSFaults}
\end{figure}

\textbf{A Cyber-Physical System (CPS)} (\citealp{alur2015principles}) comprises coupled intelligent and heterogeneous components that can cooperate, organize themselves, and make autonomous decisions (\citealp{Plateaux2016}).
The term CPS, proposed by the National Science Foundation in 2006 (\citealp{CPSNSF2018}), defines an engineered system that is built from, and depend upon, the seamless integration of computation and physical components.
CPS emphasizes the integration of computation, networking, and physical processes.
The principle challenge is an orchestration of computers (cyber) and mechanical (physical) parts.
Core CPS research areas include control, data analytics, autonomy, design, information management, Internet of Things (IoT), Human-in-the-Loop concepts (\citealp{HILLNunes15}), and networking.
CPS drives innovation in a range of dependability-critical domains, including aeronautics, energy, health care, industrial automation, and transportation.
Heterogeneous CPS components, such as sensors, computing, and network hardware, software, actuators, and physical parts are prone to various fault types, see Figure \ref{fig:CPSFaults}.
The discussed EXO-LEGS system is a typical CPS example. 

\textbf{CPS faults:} A fault is defined (\citealp{avizienis2004basic}) as an abnormal condition or a defect that may manifest itself as an error that propagates through a CPS and ultimately leads to a system failure.
For example, an internal sensor malfunction (\textit{sensor fault}) results in an erroneous measurement value (\textit{data error}) that propagates over the network to a controller where it affects the computation of a commanded actuator value (another \textit{data error}) that, in turn, leads to inability of a system to perform required function, a \textit{failure}. 
Similarly, network jitter (\textit{network fault}) results in \textit{timing errors}, the deviations in the information arrival times or signal durations (\citealp{avizienis2004basic}), that can cause control system instability (\textit{failure}).
A \textit{random fault} occurs at a random time during operation, typically due to physical processes such as damage, wear-out, or fatigue.
A \textit{systematic fault} is often a result of an error in the specification that can only be eliminated by a modification of design or the manufacturing process.
For instance, according to IEC 61508, hardware components can have both random and systematic faults.
However, software is only subject to systematic faults.
In addition, software does not degrade with time and software faults are usually input sequence dependent.
Another relevant classification of hardware faults is based on the duration that a fault is active.
Most of the hardware faults are \textit{transient faults} that appear just for a short time.
An \textit{intermittent fault} appears and disappears repeatedly, whereas a \textit{permanent fault} remains within the system if no corrective actions are performed.

\textbf{Sensor faults:} 
Sensors and sensor networks play a crucial role in CPS. 
They provide the brains of a CPS with the information about the actual physical state of the system and the environment.
In the autonomous and connected vehicle example, odometry, inertial, and ultrasonic sensors, together with more complex LIDARs, radars, and navigation cameras, provide the inputs for control, navigation, and guidance algorithms.
Similar to the most of available classifications (\citealp{jan2017sensor, reppa2016sensor, yang2016fault, dunia1996identification, kullaa2013detection, yu2015novel, yang2010fault}) we group the sensor faults according to occurred measurement errors.
\textit{Drift:} The output of the sensor keeps increasing or decreasing (usually linearly);
\textit{Offset:} The output has a (constant) difference from the correct value, also called bias;
\textit{Noise:} The variance of the sensor output significantly increases above the usual value;
\textit{stuck-at:} The output gets stuck at a fixed value, also called sensor freeze.

\textbf{Computing hardware faults:}
Computers, controllers, and embedded boards are vulnerable to Single Event Upset (SEU), a state change caused by a single ionizing particle as a result of interfering with electromagnetic fields or radiation.
A SEU can cause a bit-flip, which is an unintentional change of a bit state (\citealp{karnik2004characterization, schroeder2009dram}).
The likelihood of the occurrence of these faults is increasing due to the continuously decreasing feature sizes of integrated circuits (\citealp{de2000single, johnston2005effect, dodd2010current}).
Bit-flips are hardware transient faults that may occur in safety-critical systems, e.g. in space (\citealp{verzola2014predictive}) or automotive (\citealp{koopman2014case}) applications.
Bit-flips may affect RAM memory cells and CPU registers.
RAM stores instructions and data of the entire program, while CPU registers are temporary storages that hold the currently processing data.
Generally speaking, a CPU register loads data from memory and then processes it.
Bit-flips can be manifested in a microprocessor as e.g. timing, control-flow, and data errors (\citealp{198, svenningsson2010model}).
Bit-flips in stored loop counter values in memory cells or registers could cause timing errors.
Bit-flips in the register that stores the program counter can cause a control-flow error.
Data errors are associated with incorrect stored data values when a bit-flip alters the content of a memory cell or a register.
They are more common than timing or control-flow errors (\citealp{198}).
The following fault models have been widely, and successfully used as abstractions of physical defect mechanisms in the computing hardware: \textit{Bit stuck-at}: a permanent stuck of a bit at one (stuck-at-one) or zero (stuck-at-zero);
\textit{Random single or multiple bit-flips}: a transient random change of a bit state that can be later updated/corrected.

\textbf{Network faults:}
The CPS heterogeneity and complexity lead to major network issues, including time-varying latency (jitter), data rate limiting, packet loss, and delays (\citealp{liu2017review}).
Kremer et al. (\citealp{kremer2013cyber}) point out that real-time capability can strongly influence the design and demand in CPS application systems both from soft and hard real-time perspectives.
Ghailani et al. (\citealp{ghailanistate}) describe main issues that arise in CPS, including the network faults.
They established that practical problems with CPS are wireless network transmission errors, complex environmental interference in distributed systems, data processing errors, outliers, noise, loss of data, randomness, concreteness, data dispersion, and other uncertainties. 
Schenato et al. (\citealp{schenato2007foundations}) address shared computing and communication.
They conclude that communication delay, data loss, and time synchronization play critical roles.
Hespahna et al. (\citealp{hespanha2007survey}) present results on estimation and controller synthesis aimed at spatially distributed control systems in which the operational challenges arise from the nature of the wireless communication links between sensors, actuators, and controllers.
The paper addresses the effects of channel limitations in terms of packet rates, sampling, network delay, and packet dropouts.
Following these papers, we have identified the following network faults:
\textit{Packet loss:} one or more transmitted packets do not reach their destination;
\textit{Network delay:} one or more transmitted packets do not reach their destination within the specified time.
More complex network faults like jitter are considered to be a combination of multiple intermittent delays.

\begin{figure}[b]
  \centering
  \includegraphics[scale=0.5]{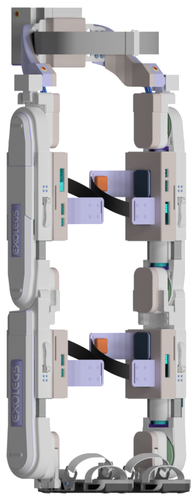}
  \caption{A CAD model of the EXO-LEGS case study.}
  \label{fig:EXO-LEGSCAD}
\end{figure}

\subsection{EXO-LEGS}
The EXO-LEGS prototype (see Figure ~\ref{fig:EXO-LEGSCAD}) has six active degrees-of-freedom in the sagittal plane.
Ankles, knees, and hips of both legs are actuated by brushless motors, whereas low-level controllers provide feedback from encoders and hall sensors and offer interfaces for the position, velocity, and current control.
The high-level controller is realized on a single-board computer connected to the joint controllers via the CAN bus using CANopen protocol.
Human motion data is captured using the off-the-shelf optical motion capture system or in-house developed system using wearable Inertial Measurement Units (IMU).
The captured motion data is preprocessed and transformed in order to generate trajectories.
The generated trajectories are used to precompute Dynamic Movement Primitives (DMP) (\citealp{dmp}).
Finally, the DMP-based controller executes precomputed primitives either in the simulation environment or on the real robot.

\section{Fault Injection Tool}
\label{sec:fit}

In this paper, we introduce a new implementation of our model-based fault injection method as a custom Simulink Fault Injection Block (FIBlock).
We used the FIBlock for the fault injection experiments in Section \ref{sec:fie}.
The presented FIBlock is available as a custom library block at (\citealp{fabarisov2020fiblock}) and implemented as a masked subsystem containing MATLAB Function block with additional Simulink Clock block as a second input.

FIBlock supports fault types that are common for the heterogeneous hardware, network, and even software components.
Thus, it can be employed for the fault injection assessment of the Cyber-Physical Systems.

\begin{figure}[b]
  \centering
  \includegraphics[width=\linewidth,scale=0.5]{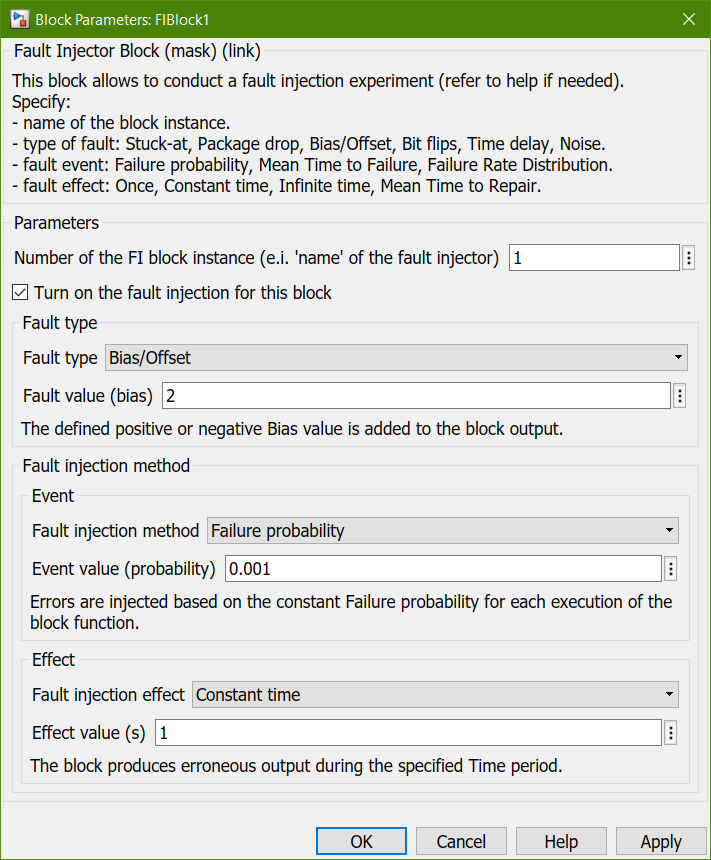}
  \caption{A GUI window of the Fault Injection Block parameters.}
  \label{fig:fibgui}
\end{figure}

The Fault Injection class is an Object-oriented programming-based implementation of the ErrorSim functions (\citealp{errorsim}).
Fault injection is performed by the instances of our Fault Injection MATLAB class.
They are being created for each FIBlock with the parameters specified by the user using the GUI of the block, see Figure  \ref{fig:fibgui}.

FIBlock allows the user to inject different types of faults.
The fault injection events and the duration of fault exposure can be modeled with different parametric stochastic methods.
The user can turn on and off a specific FIBlock, in case the fault injection in the particular block is not necessary. 
FIBlock supports the following fault types, fault events (i.e., when the fault will occur), and fault effects (i.e., the duration of the fault exposure):

\textbf{Fault types.}
\textit{Stuck-at:} the block output stays constant, preserving the latest correct value before the error occurrence;
\textit{Package drop:} the correct output is replaced by the specified value, emulating a package drop;
\textit{Bias/Offset:} the defined positive or negative Bias value is added to the block output;
\textit{Noise:} a random noise value is added. The Boundaries are defined as the percentage of the correct value;
\textit{Time delay:} a delay is introduced into the signal. During the delay the value is the same as it was before the fault activation;
\textit{Bit flips:} the defined Number of bits are inverted in the binary representation of the correct value.

\textbf{Fault events.}
\textit{Failure probability:} errors are injected based on the constant failure probability for each execution of the block function;
\textit{Mean Time to Failure:} errors are injected according to the specified MTTF. Normal distribution.

\textbf{Fault effects.}
\textit{Once:} an error appears only one time during a simulation;
\textit{Constant time:} the block produces erroneous output during the specified time period;
\textit{Infinite time:} the block produces erroneous output until the end of the simulation;
\textit{Mean Time to Repair:} normally distributed MTTR regulates the time of the error effect.

\begin{figure}[h]
  \centering
  \includegraphics[width=\linewidth]{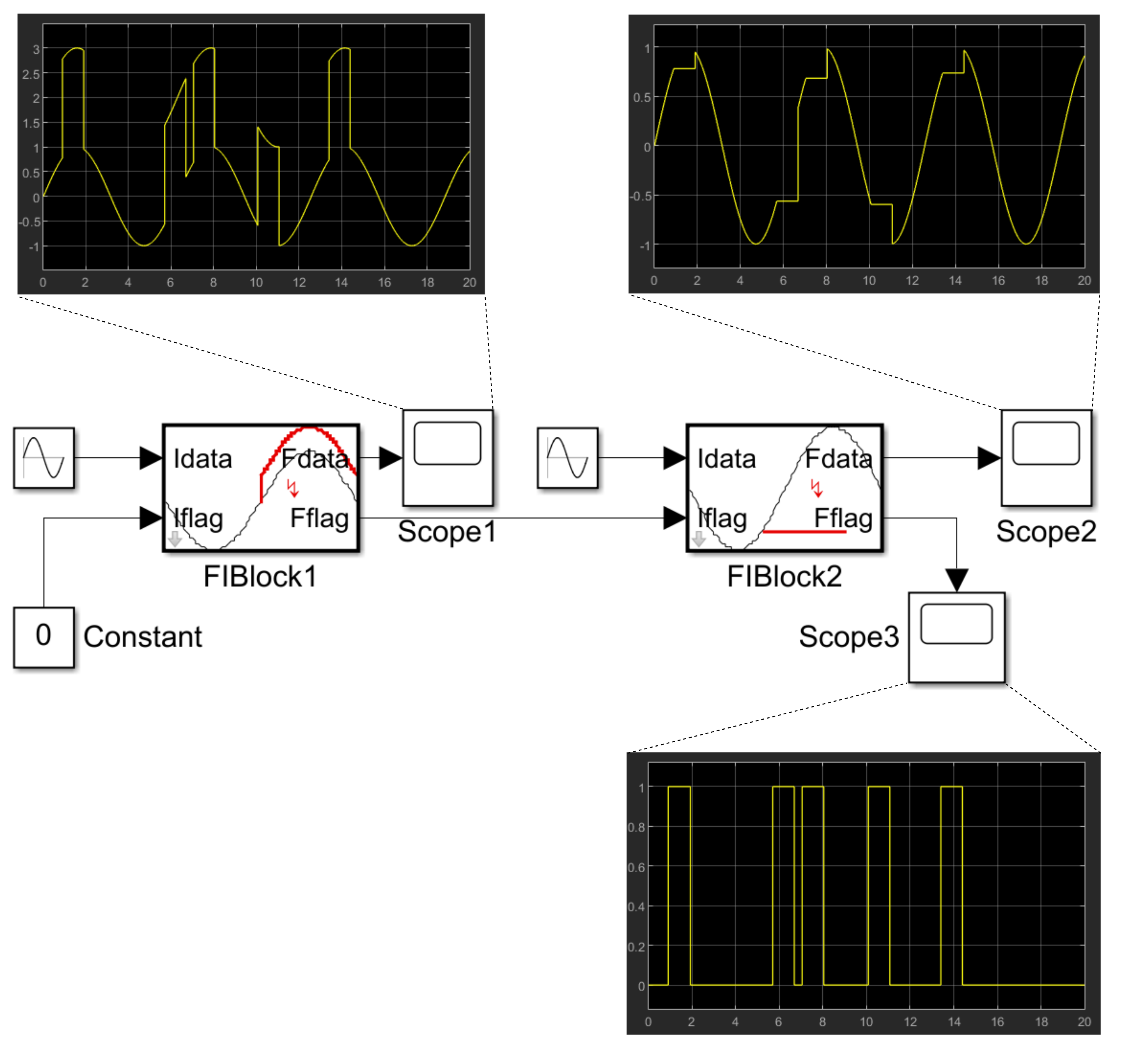}
  \caption{An example of a Simulink system with two connected FIBlocks.}
  \label{fig:fibexample}
\end{figure}

The injection of faults in multiple locations is called \textit{multiple-point fault} injection (\citealp{ayatolahi2013study}).
The situation when the activation of one fault triggers the activation of another fault is called \textit{conditional} or \textit{chained faults}.
FIBlock supports both multiple-point and chained faults.
Figure \ref{fig:fibexample} shows a Simulink model with two FIBlocks.
A second output of the first block emits an error trigger signal.
This output signal is connected to the second input of the second FIBlock.
In the case of the fault activation in the first block, the emitted trigger signal would force the fault activation in the second block.

Also, MATLAB Function blocks support code generation with Simulink Coder, therefore it is possible to generate code for the controller with the FIBlocks.
This allows the fault injection not only in the Simulink environment but in the real hardware.

\section{EXO-LEGS Simulink model}
\label{sec:cs}

The exoskeleton Simulink model consists of seven rigid bodies and six joints.
All the joints are modeled using Simscape Multibody blocks.
The leg model consists of four blocks, namely the waist, the hip-knee component, the knee-ankle component, and the foot component.
The DMP Block controls the dynamic stability of the exoskeleton along the desired trajectory.
The stepDMPBlock is in charge of executing the DMPs.
DMP is a machine learning compatible method of modeling the attractor behavior of nonlinear dynamical systems (\citealp{dmp}).
A DMP system is a collection of multiple individual DMPs.
Each DMP controls one joint.
The DMPs are kept in sync by a single central Canonical System that manages the time for the complete DMP system.
The DMPs were developed by modulating a damped spring model with nonlinear terms to get a trajectory control method:

\begin{equation}
\tau \dot{z} = \alpha_z (\beta_z (g-y)-z)+f
\end{equation} 

\begin{equation}
z = \tau \dot{y}
\end{equation} 

\noindent where $\ddot{y}, \dot{y}$, and $y$ are desired acceleration, velocity and position of the system.
$\alpha_z$ and $\beta_z $ are damping parameters and can be made critically damped by choosing them as $\beta_z=\alpha_z /4$ to converge against the goal $g$.
$\tau$ is the temporal scaling factor and determines the relative passage of time in the system to the real time.
Increasing $\tau$ results in a slow down, thereby a decrease of $\tau$ will speed the canonical system up.
$f$ is the forcing term that holds the trajectory information but if chosen as $f=0$ "represents a globally stable second-order linear system with (z,y)=(0,g) as a unique point attractor" (\citealp{dmp}).
Finally, the dynamic\textunderscore control block receives current velocities and positions either from the simulation or from the real robot and calculated by stepDMPBlock target accelerations, velocities, and positions in order to compute target torques for each joint.

\section{Fault injection experiments}
\label{sec:fie}

We use the FIBlock, presented in Section \ref{sec:fit}, to inject typical sensor faults and bit-flips into the Simulink model of EXO-LEGS, discussed in Section \ref{sec:cs}. 
The device has different specifications of actuators due to the different requirements of torque and speed of the joints.
These joints assist the user with 30\% of the power needed to walk.

Joint torque and speed requirements for the straight walking movement are presented in the Table \ref{tab:constr}.
Power joints used in EXO-LEGS were chosen based on the following equation:

\begin{equation}
  P_j = T_j \times \omega = T_j \times \frac{2\pi}{60} \times n_j
\end{equation}

\noindent where \(P_j\) is power at the joint j [W]; \(T_j\) is torque at the joint j [N.m]; \(\omega\) is angular velocity [rad/sec]; \(n_j\) is speed velocity at the joint j [RPM].
Because power joints were chosen based on (1), they ought to withhold the torque and speed no greater than the one presented in the Table \ref{tab:constr}.
Other constraints that need to be accounted for are range of travel (ROT) of the exoskeleton joints. 
These mechanical stop limits restraint flexion and extension movements in the hip, knee, and ankle according to the following values:
\textit{hip}: -30º to 90º; \textit{knee}: -90º to 0º; \textit{ankle}: -30º to 30º.
In the experiment we consider the situation, when the angular position of a joint is out of ROT as a system failure. Since the violation of the biomechanical joint limits could lead to critical injuries of the user.

According to (\citealp{laprie1992dependability}), a \textit{fault} is a defect in the system that can be activated and cause an error.
An \textit{error} is an incorrect internal state of the system, a discrepancy between the intended behavior of a system and its observed behavior.
Such an invalid internal system state generated by an error may lead to another error or failure.
A \textit{failure} is an instance in time when the system displays behavior that is contrary to its specification.
According to the aforementioned constraints and requirements, we can specify the threshold for what to consider as errors and failures of the system, see Table \ref{tab:constr}.

\begin{table}
\tbl{Torque and speed requirements for correct straight walking.}
{\label{tab:constr}
\tabcolsep 3pt
\begin{tabular}{|l|l|l|l|l|}
\hline
                              & \textbf{Hip} & \textbf{Knee} & \textbf{Ankle} & \textbf{Error/Failure} \\ \hline
\textbf{Max. torque {[}Nm{]}} & 72.9         & 54.9          & 128.7          & Error                  \\ \hline
\textbf{Max. speed {[}rpm{]}} & 23.4         & 65.2          & 50.8           & Error                  \\ \hline
\textbf{Min. angle {[}deg{]}} & -30          & -90           & -30            & Failure                \\ \hline
\textbf{Max. angle {[}deg{]}} & 90           & 0             & -30            & Failure                \\ \hline
\end{tabular}}
\end{table}

\begin{figure}[b]
  \centering
  \includegraphics[width=\linewidth]{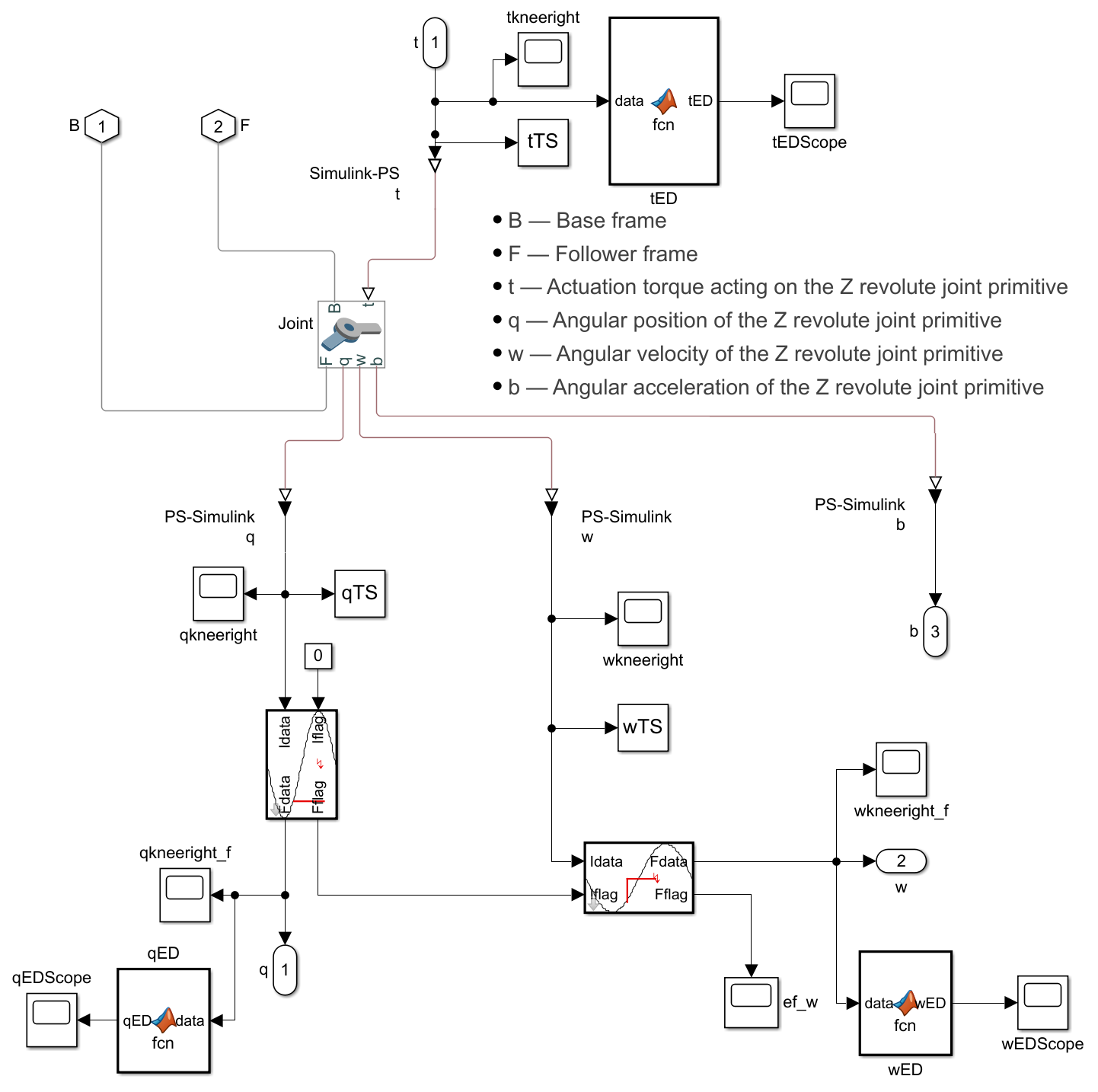}
  \caption{A subsystem of the case study model representing the right knee with FIBlocks.}
  \label{fig:fibmodel}
\end{figure}

\textbf{Fault injection scenario:}
The motion that has been used for evaluation is straight walking.
Based on the state of the art review provided in Section \ref{sec:sta}, the following fault injection scenario has been selected.
With the probability $0.0005$, the \textit{'Stuck-at current value'} fault is injected into the angular position signal.
The duration of the fault effect is varying in time.
Using the chained fault injection mechanism the \textit{'Stuck-at 0'} (also referred as sensor freeze) is injected into the angular velocity signal whenever the previously mentioned fault is activated.
This fault is activated with the same duration as the previous one.
The probability of the fault activation was dictated by the fact that with the greater probability, the faults tend to occur too frequently, with a little to no time interval between them.
This would often lead to an immediate system failure and didn’t allow us to comprehensively compute the reliability metrics.
As discussed in the following paragraph, when the faults occur one after another, the controller tries to overcompensate them, leading to system failures (Figure \ref{fig:fibmodel}).
On the other hand, the lower probability frequently leads to no faults during the seven seconds of the simulation.

\textbf{Goals of the experiments:} The first goal of these experiments was to determine the critical fault duration at which the controller could no longer mitigate the errors.
The second goal was to ensure that the exoskeleton would not fall with the angular degree exceeding the constraint pins for the cases when the controller is able to mitigate the errors, and the correct movement trajectory is preserved.

\begin{figure}[b]
  \centering
  \includegraphics[width=\linewidth]{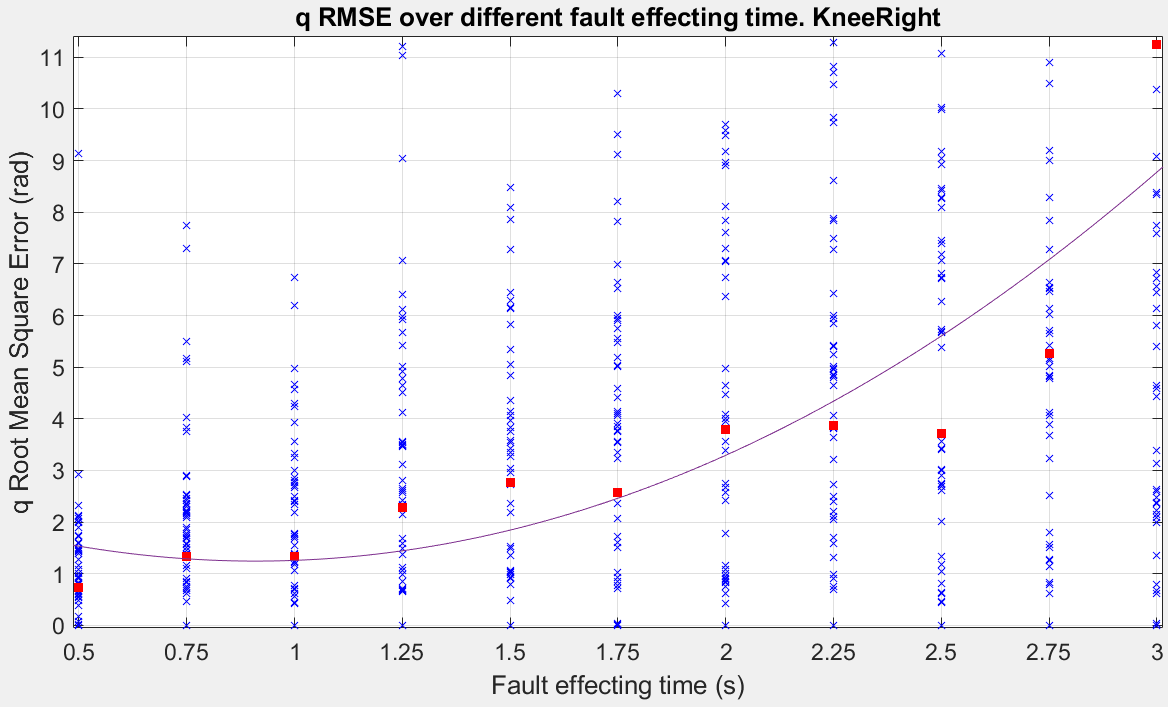}
  \caption{RMSE of the angular position signal. Fault duration varies from 0.5s to 3s with a 0.25s step.}
  \label{fig:qrmpseresult}
\end{figure}

\begin{figure}[b]
  \centering
  \includegraphics[width=\linewidth]{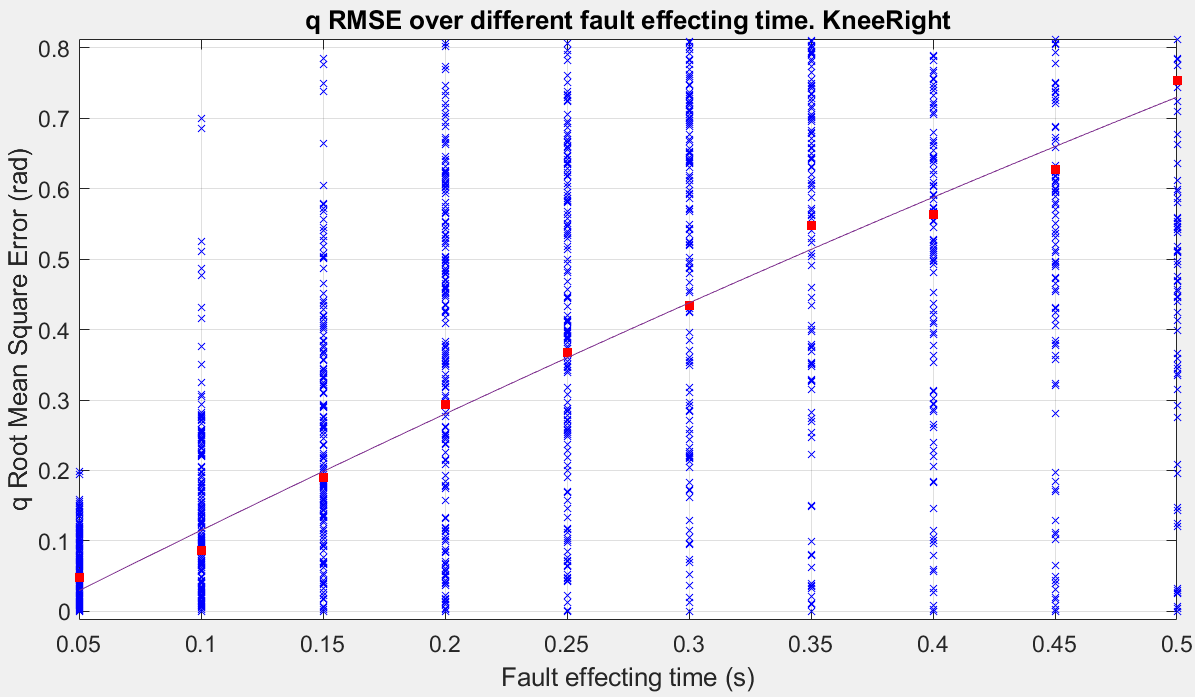}
  \caption{RMSE of the angular position signal. Fault duration varies from 0.05s to 0.5s with a 0.05s step.}
  \label{fig:qrmseresult}
\end{figure}

There are two sets of fault injection experiments.
First, we vary the fault duration (fault effect) from 0.5s to 3s with the 0.25s step, with the fault being injected in the right knee joint.
We compute the Root Mean Square Error (RMSE) of the following sensor signals: \textit{angular position}, \textit{angular velocity}, and \textit{actuation torque} acting on joint.
Figure \ref{fig:qrmpseresult} presents calculated RMSE values at each fault effect iteration.
The red squares denote mean values.
The purple spline is the quadratic fitting curve.
These results show that even with the fault duration of 1 second, the difference between the intended trajectory and the faulty one is already too big (1.337 rad).
In the second set of experiments, we vary the fault duration from 0.05s to 0.5s with the step equal to 0.05s.
The results are presented in the Figure \ref{fig:qrmseresult}.
The intended trajectory deviation varies from 0.047 rad to 0.754 rad respectively.

\textbf{Evaluation of the experiment results:}
The experiments helped us to conclude that the system can successfully compensate injected faults if they are less than 0.1 sec.
Any fault with a longer duration will lead to a critical injury of the patient (i.e., system failure defined in the Table \ref{tab:constr}).
However, such a fault wouldn't necessarily lead to an error defined in the Table \ref{tab:constr} (e.g., in case if they occurred with a time interval between them).

\textbf{An additional set of fault injection experiments} was aimed to the evaluation of the bit-flips.
According to  (\citealp{lowery2013relative}), a bit-flip, being a single event upset, would rather often lead to insignificant error in floating-point value signal.
The fault injection experiment results corroborated such a theory.
The same result was obtained with the injection of single spikes and offsets.
The controller successfully compensated such fault types.

\section{Conclusion}
In this paper, we presented our new model-based fault injection method implemented as a custom Simulink Fault Injection Block (FIBlock).
It's available as an open-source tool on the GitHub.
FIBlock allows the injection of different types of faults, such as stuck-at, package drop, bias/offset, bit flips, time delay, and noise.
Such a fault injector, implemented as a highly-customizable Simulink block library, is easy to integrate into a Simulink model.
FIBlock supports both dual-point and chained fault injection.
We used the FIBlock for the fault injection experiments for the assessment of the fault tolerance of an exoskeleton system.

During the fault injection experiments, we were injecting simultaneously the \textit{'Stuck-at current value'} fault into the angular position signal, and the \textit{’Stuck-at  0’} fault into the angular velocity signal.
The probability in every case was $0.0005$, while the duration was variable.
Upon the evaluation of the fault injection experiments, we discovered that the ability of the controller to compensate for the error values mostly depends not only on a fault duration but also on the fault occurrence frequency.
In case of non-consecutive faults, the controller can effectively compensate fault duration not greater than 0.3s.
Effectively in this case means that it didn't lead to failure, although errors were nevertheless arising.
However, in the case of consecutive faults, the controller's attempts to compensate them were leading to a system failure.
With a fault duration not greater than 0.1 seconds, the controller was able to compensate it.
In future, we will perform fault injection experiments on a physical exoskeleton system to acquire more insights about the application of the proposed method in the real-world environment.

\bibliographystyle{Chicago}
\bibliography{bibls}

\end{document}